\documentclass[aps,prl,twocolumn,superscriptaddress,showpacs]{revtex4-1}
\usepackage{graphicx} %For including figures

\begin{document}

\title{Shubnikov-de Haas oscillations of high mobility holes in monolayer and bilayer WSe$_2$: Landau level degeneracy, effective mass, and negative compressibility}

%Author Names and Affiliations
\author{Babak Fallahazad}
\author{Hema C. P. Movva}
\author{Kyounghwan Kim}
\author{Stefano Larentis}
\affiliation{Microelectronics Research Center, Department of Electrical and Computer Engineering, The University of Texas at Austin, Austin, Texas 78758, USA}
\author{Takashi Taniguchi}
\author{Kenji Watanabe}
\affiliation{National Institute of Materials Science, 1-1 Namiki Tsukuba, Ibaraki 305-0044, Japan}
\author{Sanjay K. Banerjee}
\author{Emanuel Tutuc}
\email{etutuc@mer.utexas.edu}
\affiliation{Microelectronics Research Center, Department of Electrical and Computer Engineering, The University of Texas at Austin, Austin, Texas 78758, USA}

\date{\today}

\begin{abstract}
We study the magnetotransport properties of high mobility holes in monolayer and bilayer WSe$_2$, which display well defined Shubnikov-de Haas (SdH) oscillations, and quantum Hall states (QHSs) in high magnetic fields. In both mono and bilayer WSe$_2$, the SdH oscillations and the QHSs occur predominantly at even filling factors, evincing a two-fold Landau level degeneracy. The Fourier transform analysis of the SdH oscillations in bilayer WSe$_2$ reveal the presence of two subbands localized in the top or the bottom layer, as well as negative compressibility.  From the temperature dependence of the SdH oscillations we determine a hole effective mass of $0.45m_{0}$ for both mono and bilayer WSe$_2$.
\end{abstract}


\maketitle

% Main text
Molybdenum and tungsten-based transition metal dichalcogenides (TMDs) in the 2H phase are characterized by a honeycomb lattice similar to graphene. The conduction and valence band minima in monolayer TMDs are reached at the corners ($K$ points) of the first Brillouin zone, and away from the band minima the broken inversion symmetry combined with the strong spin-orbit coupling lifts the four-fold (spin-valley) degeneracy, and yields coupled spin and valley degrees of freedom \cite{xiao_coupled_2012}.
% States with the same momentum and opposite spins in the same valley are no longer degenerate, but states with opposite spins in opposite valley are degenerate, thanks to time reversal invariance.
The spin-valley coupling present in TMDs has been probed extensively using optical excitation, thanks to peculiar selection rules \cite{xiao_coupled_2012, mak_control_2012, jones_optical_2013, li_valley_2014, jones_spin-layer_2014}.  In perpendicular magnetic fields the spin-valley coupling translates into two-fold degenerate Landau levels (LLs) in TMDs \cite{li_unconventional_2013,Rose_PRB2013}, as opposed to the case of graphene where single particle states in LLs are four-fold degenerate \cite{Novoselov22102004, zhang_experimental_2005}.  Exploring the TMD electron physics at low temperatures and high magnetic fields has proven more arduous because of the moderate mobility, combined with the high resistance of metal-TMD contacts at reduced temperatures.  Attempts to address this issue include using contacts such as graphene on MoS$_2$ \cite{cui_multi-terminal_2015}, or Pt underneath WSe$_2$ \cite{movva_high-mobility_2015}. To reduce the surface roughness, and charged impurity scattering in ultra-thin TMDs, atomically flat dielectrics such as hexagonal boron-nitride (hBN) are preferable \cite{deanc._r._boron_2010}.

Here we present a magnetotransport study of dual-gated mono and bilayer WSe$_2$ with top and bottom hBN gate dielectrics, and bottom Pt contacts. Both mono and bilayer WSe$_2$ samples exhibit Shubnikov-de Haas (SdH) oscillations in perpendicular magnetic fields with a density-to-frequency ratio of $2e/h$, indicating a two-fold LL degeneracy; $e$ is the electron charge, and $h$ Planck's constant.  In bilayer WSe$_2$, we observe quantum Hall states (QHSs) at even filling factors $\nu= 6, 8, 10,...,$ which further confirm the two-fold degenerate LLs.  At the highest magnetic field we observe a $\nu=5$ QHS which signals a full lifting of the LL degeneracy.  The Fourier transform (FT) analysis on the SdH oscillations in bilayer WSe$_2$ reveals the presence of two subbands, each localized in the top or the bottom layer, as well as negative compressibility.  Using the SdH temperature dependence we determine an effective hole mass in mono and bilayer WSe$_2$ of $m^{*} = 0.45m_{0}$, where $m_0$ is the bare electron mass.

\begin{figure*}
\centering
\includegraphics[scale=0.65]{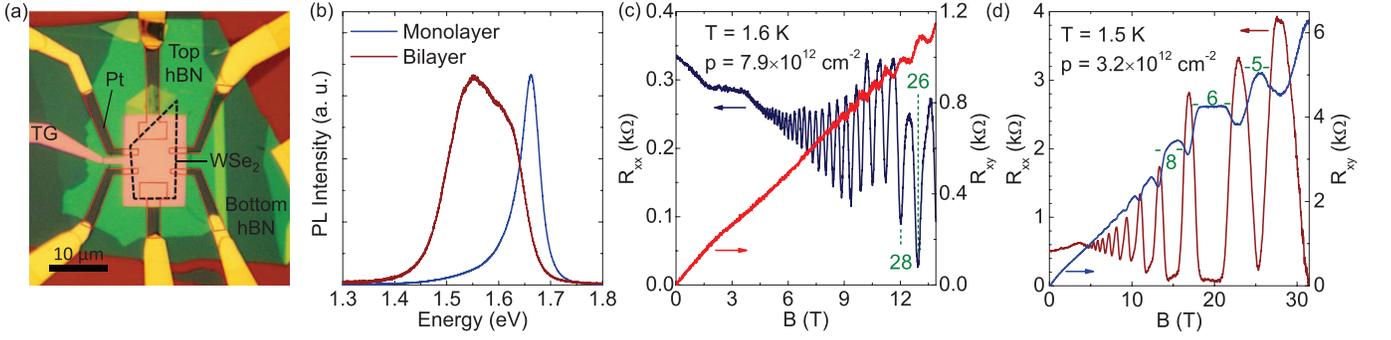}
\caption {\small{(a) Optical micrograph of a dual-gated WSe$_2$ sample with bottom Pt contacts. The dashed black (red) contour marks the boundaries of the WSe$_2$ flake (Pt contacts).
(b) PL spectra of mono and bilayer WSe$_2$ normalized to the highest intensity. (c) Monolayer WSe$_2$ $R_{xx}$ (left axis) and $R_{xy}$ (right axis) vs. $B$ measured at $T$ = 1.6 K, at a hole density
$p = 7.9\times10^{12}$ cm$^{-2}$.  The two lowest filling factors, $\nu = 26$ and $\nu= 28$ are indicated. (d) Bilayer WSe$_2$ $R_{xx}$ (left axis) and $R_{xy}$ (right axis) vs. $B$ measured at $T$ = 1.5 K,
and at $p = 3.2\times10^{12}$ cm$^{-2}$. A quantized $R_{xy}$ plateau is observed at $\nu = 6$.}}
\label{fig1}
\end{figure*}

Figure~\ref{fig1}(a) shows an optical micrograph of a dual gated WSe$_2$ Hall bar sample encapsulated by hBN dielectrics, and fabricated using a layer pick-up method similar to Ref. \cite{movva_high-mobility_2015}. The samples use synthetic WSe$_2$ (HQGraphene, CAS number: 12067-46-8) mechanically exfoliated on SiO$_2$/Si substrates. Using a combination of optical contrast, Raman spectroscopy, and photoluminescence (PL) spectroscopy, mono and bilayer WSe$_2$ flakes are identified. Thanks to the thickness dependence of the band structure, mono and bilayer WSe$_2$ possess distinct PL signatures that unambiguously differentiate them from thicker WSe$_2$ \cite{terrones_new_2014, kim_band_2015, zhao_evolution_2013}. Figure~\ref{fig1}(b) shows sample PL spectra of mono and bilayer WSe$_2$ as exfoliated, measured at an incident excitation wavelength of 532 nm. Monolayer WSe$_2$ shows a single peak at 1.65 eV, consistent with a direct energy gap, and in good agreement with previously reported energy gap values \cite{terrones_new_2014, zhao_evolution_2013}. Bilayer WSe$_2$ shows a broader peak that can be fitted with two Lorentzian peaks centered at 1.55 eV and 1.61 eV, reflecting the transition to indirect energy gap \cite{terrones_new_2014, zhao_evolution_2013, liu_electronic_2015}. The WSe$_2$ flakes were also investigated using Raman spectroscopy, where a distinct difference between the spectra of mono and bilayer WSe$_2$ is the presence of the $A_{1g}^{2}$ mode at 310 cm$^{-1}$ \cite{terrones_new_2014}. Using mechanically exfoliated hBN flakes combined with micromanipulation and transfer techniques, dual-gated WSe$_2$ samples encapsulated in hBN dielectrics, and with bottom Pt contacts \cite{movva_high-mobility_2015} are fabricated.  Four dual-gated WSe$_2$ samples, two monolayers and two bilayers, were investigated in this study, all with consistent results.  Here we focus on data from two samples, one monolayer WSe$_2$ and one bilayer WSe$_2$. The samples are characterized using small signal, low frequency lock-in techniques, at temperatures down to $T = 1.5$ K, and perpendicular magnetic fields up to $B = 31.5$ T.

Examples of longitudinal ($R_{xx}$) and Hall ($R_{xy}$) resistance measured as a function of the $B$-field at fixed carrier densities for mono and bilayer WSe$_2$ are shown in Fig.~\ref{fig1}(c) and Fig.~\ref{fig1}(d), respectively. The $R_{xx}$ vs. $B$ data of Fig.~\ref{fig1}(c), measured at $T$ = 1.6 K, top gate voltage $V_{TG}$ = $- 6$ V, and back gate voltage $V_{BG}$ = 0 V shows well-defined SdH oscillations starting at $B$ $\cong$ 4.5 T.  The filling factors corresponding to the two lowest LLs probed in this measurement, $\nu$ = 26 and $\nu$ = 28, are marked. The hole density ($p$) calculated from the slope of $R_{xy}$ vs. $B$ at low fields is $p$ = $7.9\times10^{12}$ cm$^{-2}$.

\begin{figure*}
\centering
\includegraphics[scale=0.55]{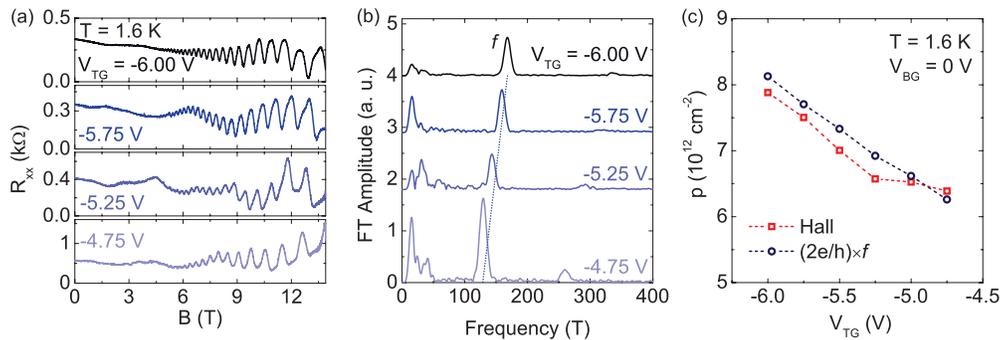}
\caption {\small{(a) $R_{xx}$ vs. $B$ measured in monolayer WSe$_2$ at different $V_{TG}$ values, $V_{BG}$ = 0 V, and $T$ = 1.6 K. (b) FT amplitude vs. frequency corresponding to $R_{xx}$ vs. $B^{-1}$ data of panel (a) data; the traces are shifted for clarity. The dashed line is a guide to the eye. (c) $p$ vs. $V_{TG}$ of monolayer WSe$_2$ measured at $V_{BG}$ = 0 V and $T$ = 1.6 K. The red symbols (rectangles) represent the Hall density, and the blue symbols (circles) show $p$ = $(2e/h)\times{f}$. The agreement confirms the two-fold LL degeneracy.}}
\label{fig2}
\end{figure*}

The bilayer WSe$_2$ magnetotransport data of Fig.~\ref{fig1}(d) are measured at $T$ = 1.5 K, $V_{TG}$ = $-6.4$ V and $V_{BG}$ = $60$ V. The hole density is $p$ = $3.2\times10^{12}$ cm$^{-2}$. Similar to the monolayer WSe$_2$ case, the SdH oscillations are present in bilayer WSe$_2$, along with developing QHSs at even filling factors. The data show developed QHSs accompanied by $R_{xy}$ plateau at $\nu$ = 6 and $\nu$ = 8, along with an onset of the $\nu$ = 4 QHS at $B$-fields larger than 31 T.  Furthermore, a developing QHS at $\nu$ = 5 signals a full lifting of the LL degeneracy. Figures~\ref{fig1}(c) and~\ref{fig1}(d) data combined suggest the oscillations predominantly occur at even filling factors in both mono and bilayer WSe$_2$.

Figure~\ref{fig2}(a) shows $R_{xx}$ and $R_{xy}$ vs. $B$ data measured in monolayer WSe$_2$ at different $V_{TG}$ values, $V_{BG}$ = 0 V, and $T$ = 1.6 K. Figure~\ref{fig2}(b) shows the FT amplitude vs. frequency corresponding to the $R_{xx}$ vs. $B^{-1}$ of Fig.~\ref{fig2}(a).  The FT is calculated by first subtracting a third order polynomial from the $R_{xx}$ vs. $B^{-1}$ data to center the oscillations around zero, multiplying the data by a Hamming window, and finally applying a fast Fourier transform algorithm. The FT data reveals a principal peak at a frequency ($f$), along a smaller amplitude  second harmonic ($2f$). The $f$ value increases with $|V_{TG}|$, and therefore with increasing the hole density.  For a 2D carrier system the SdH frequency-density dependence is $f = 1/g\times(h/e)\times{p}$, where $g$ is the LL degeneracy.  For example, $g$ = 1 ($g$ = 2) for spin resolved (degenerate) LLs, or $g$ = 4 for spin and valley degenerate LLs, as in the case for Si \cite{fowler_magneto-oscillatory_1966}, AlAs \cite{shkolnikov_valley_2002}, and graphene 2D systems \cite{Novoselov22102004, zhang_experimental_2005}.  To determine the LL degeneracy factor in monolayer WSe$_2$, we examine the ratio of $f$ to the carrier density determined from Hall measurements. Figure~\ref{fig2}(c) shows $p$ vs. $V_{TG}$ measured at $V_{BG}$ = 0 V, and $T$ = 1.6 K in monolayer WSe$_2$.  The carrier density values determined from the FT analysis using $p = (2e/h)\times{f}$ are included for comparison, and the agreement confirms the $g$ = 2 LL degeneracy.

\begin{figure*}
\centering
\includegraphics[scale=0.56]{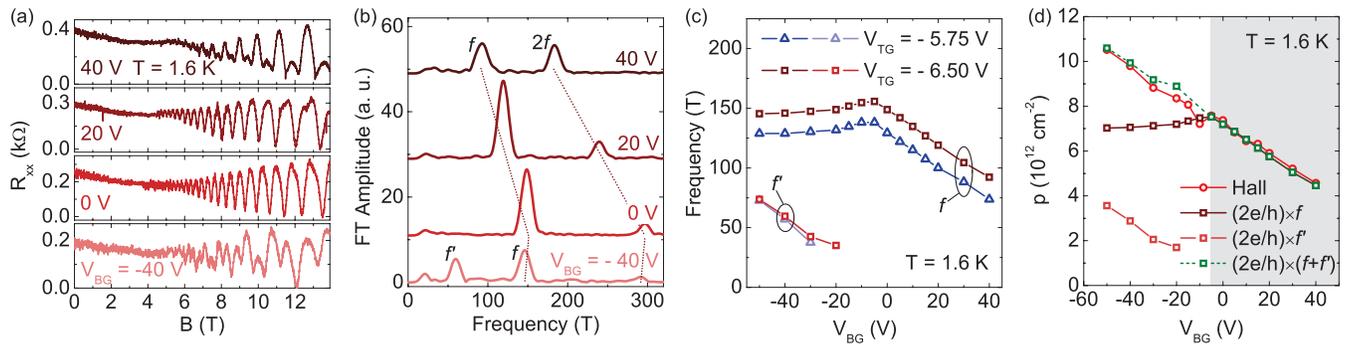}
\caption {\small{(a) Bilayer WSe$_2$ $R_{xx}$ vs. $B$ measured at various $V_{BG}$ values, $V_{TG}$ = $-6.5$ V, and $T$ = 1.6 K. (b) FT amplitude vs. frequency of panel (a) $R_{xx}$ vs. $B^{-1}$ data; the traces are shifted for clarity. At negative $V_{BG}$ an additional peak ($f'$) emerges, and concomitantly $f$ no longer increases with $V_{BG}$. The dashed lines are a guide to the eye. (c) $f$ and $f'$ vs. $V_{BG}$ in bilayer WSe$_2$ at $V_{TG}$ = $-5.75$ V (triangles),  $V_{TG}$ = $-6.50$ V (rectangles). (d) $p$ vs. $V_{BG}$ of bilayer WSe$_2$ measured at $V_{TG}$ = $-6.5$ V. The dark (light) red squares mark the top (bottom) layer hole density. The shaded area marks the bottom layer depopulation.}}
\label{fig3}
\end{figure*}

Figure~\ref{fig3}(a) shows $R_{xx}$ vs. $B$ measured in bilayer WSe$_2$ at different $V_{BG}$ values, $V_{TG}$ = $-6.5$ V, and $T$ = 1.6 K.  The data show SdH oscillations with a beating pattern at negative $V_{BG}$.  Figure~\ref{fig3}(b) shows FT of $R_{xx}$ vs. $B^{-1}$ of Fig.~\ref{fig3}(a) data at different $V_{BG}$ values. The FT data at positive $V_{BG}$ possess a main peak at a frequency $f$ along with its second harmonic ($2f$) consistent with the monolayer WSe$_2$ Fig.~\ref{fig2}(b) data. Figure~\ref{fig3}(b) data show that at $V_{BG}$ = $-40$ V an additional peak emerges at a lower frequency ($f'$). The additional peak ($f'$) is absent in monolayer WSe$_2$. Figure ~\ref{fig3}(c) summarizes the $f$ and $f'$ frequency values vs. $V_{BG}$ in bilayer WSe$_2$ at two different $V_{TG}$ values, and at $T$ = 1.6 K. There are several noteworthy features of Fig.~\ref{fig3}(c) data.  First, both $f$ and $f'$ have a linear dependence on $V_{BG}$, albeit in different ranges, positive (negative) $V_{BG}$ for $f$ ($f'$). Second, the emergence of the additional peak ($f'$) at negative $V_{BG}$ coincides with $f$ becoming weakly dependent on $V_{BG}$.  Third, at a fixed $V_{BG}$ the value of $f$ increases with $|V_{TG}|$, suggesting that $f$ responds to the carrier density induced by the top gate. When present, $f'$ is insensitive to $V_{TG}$, but depends linearly on $V_{BG}$, suggesting that $f'$ responds to the carrier density induced by the back gate.  The combined $V_{BG}$ and $V_{TG}$ dependence suggests the peak at frequency $f$ is determined by the hole density induced in the top layer, while the peak at $f'$ is associated with the hole density in the bottom layer of the bilayer WSe$_2$.  At $V_{BG}$ $>$ 0 V, the bottom layer in bilayer WSe$_2$ is fully depleted, while the top layer is populated by the applied $V_{TG}$.  A negative $V_{BG}$ populates the bottom layer.  When both layers are populated the $f$ and $f'$ frequencies respond largely to the applied $V_{TG}$ and $V_{BG}$, respectively, and are insensitive to the opposite gate thanks to screening. To verify the above interpretation, Fig.~\ref{fig3}(d) shows a comparison of $p$ vs. $V_{BG}$ determined from Hall measurements, and from the SdH oscillations, namely $p = (2e/h)\times(f+f')$. The agreement confirms that $f$ and $f'$ are determined by the top, and bottom layer densities, respectively.

\begin{figure*}
\centering
\includegraphics[scale=0.57]{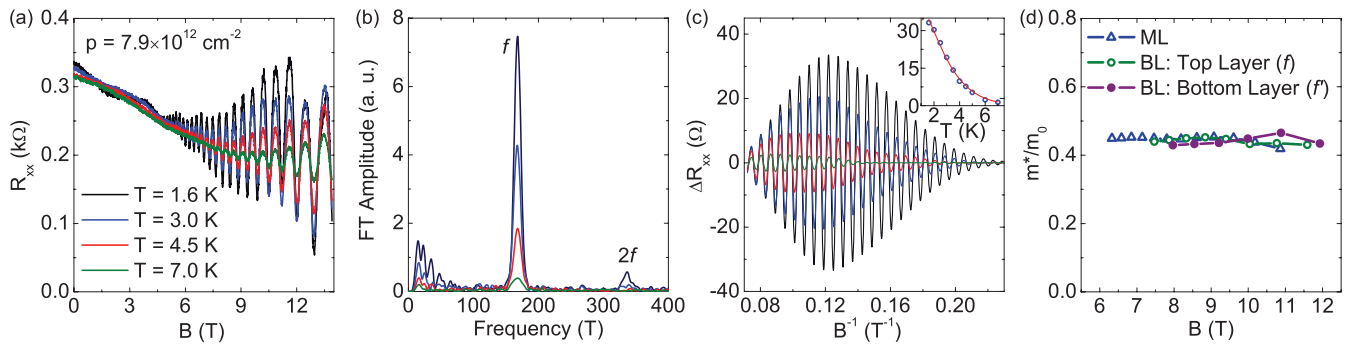}
\caption {\small{(a) $R_{xx}$ vs. $B$ in monolayer measured at different $T$ values. (b) FT amplitude vs. frequency corresponding to panel (a) $R_{xx}$ vs. $B^{-1}$ data. (c) $\Delta{R_{xx}}$ vs. $B^{-1}$ calculated from the inverse FT of panel (b) data, using a band pass filter to isolate the peak $f$. Inset: $\Delta{R_{xx}}$ vs. $T$ measured at $B$ = 7.81 T (symbols), along with the Dingle factor fit (red line). (d) $m^{*}/m_{0}$ vs. $B$ in monolayer (ML), and bilayer (BL) WSe$_2$.  The bilayer data includes values for both peaks (layers).}}
\label{fig4}
\end{figure*}

Figure ~\ref{fig3}(c,d) data are similar to the layer density dependence on gate bias in GaAs double quantum wells separated by a tunneling barrier \cite{katayama_charge_1995}.  Moreover, a peculiar feature of Fig.~\ref{fig3}(c,d) data is that the frequency $f$, and therefore the top layer density decreases with increasing the total density once the bottom layer becomes populated, signaling negative compressibility for the holes in the bottom WSe$_2$ layer as a result of interaction.  Similar observations have been reported for 2D electrons in GaAs \cite{eisenstein_compressibility_1994}, LaAlO$_3$/SrTiO$_3$ heterojunctions \cite{li_very_2011}, MoS$_2$ \cite{larentis_band_2014}, and more recently bulk WSe$_2$ \cite{riley_negative_2015}.

We turn now to the effective mass measurements. The SdH oscillation amplitude (${\Delta}R_{xx}$) is proportional to the Dingle factor, ${\xi}$/sinh ${\xi}$, where $\xi  = \frac{2{\pi}^{2}k_{B} T}{\hbar{\omega_{c}}}$, and $\omega_{c}=\frac{eB}{m^{*}}$ is the cyclotron frequency \cite{dingle_magnetic_1952, adams_quantum_1959}; $\hbar$ is the reduced Planck constant, and $k_B$ is Boltzmann constant.  Figure~\ref{fig4}(a) shows the $R_{xx}$ vs. $B$ data measured in monolayer WSe$_2$ at $p$ = $7.9\times10^{12}$ cm$^{-2}$, and at temperatures ranging between $T$ = 1.6 K and 7.0 K. Figure~\ref{fig4}(b) presents the FT associated with Fig.~\ref{fig4}(a) data. Data similar to Fig. 4(a) and 4(b) were acquired for bilayer WSe$_2$. We use the following procedure to extract $m^*$, which can be applied to both a single subband (e.g. monolayer WSe$_2$) or a multisubband 2D system (e.g. bilayer WSe$_2$).  At a fixed temperature, we first obtain the FT of $R_{xx}$ vs. $B^{-1}$ data as discussed in Fig.~\ref{fig2}.  We then apply a band pass filter centered at $f$, or $f'$ for the case of bilayer WSe$_2$, and perform an inverse FT.  Figure~\ref{fig4}(c) shows $\Delta{R_{xx}}$ vs. $B^{-1}$ at different $T$ values, obtained via the inverse FT from Fig.~\ref{fig4}(b) data.  At a fixed $B$-field, the $\Delta{R_{xx}}$ vs. $T$ data of Fig.~\ref{fig4}(c) is fitted to the Dingle factor to obtain $m^{*}$ [Fig.~\ref{fig4}(c) inset].  Figure~\ref{fig4}(d) summarizes the extracted $m^{*}$ values vs. $B$ for both mono and bilayer WSe$_2$. In monolayer, and also in either subband of the bilayer WSe$_2$ $m^{*}=0.45m_0$, independent of the $B$-field. The reported theoretical $m^{*}/m_{0}$ values for the upper valence band in monolayer WSe$_2$ include 0.33 \cite{li_unconventional_2013}, 0.34 \cite{shi_quasiparticle_2013}, 0.36 \cite{kormanyos_k_2015, zibouche_electron_2014}, 0.43 \cite{fang_textitab-initio_2015}, and 0.46 \cite{chang_ballistic_2014}.

Lastly, we expand on the observed QHSs sequence. The valence band LLs cyclotron energies are $E_{n}=-n\hbar\omega_{c}$; $n$ is the LL index. The LL degeneracy is 1 for $n=0$, and 2 for $n>0$, leading to expected QHSs at odd filling factors \cite{li_unconventional_2013,Rose_PRB2013}.  The experimental data shows a two-fold LL degeneracy, but the QHSs occur at predominantly even fillings. This can be explained by considering the LL valley and spin Zeeman energy $E_{\tau s}=g_{v}\tau\mu_B B+ g_{e} s \mu_B B$, where $\mu_B$ is the Bohr magneton, $\tau=\pm1$ corresponds to the $K$ and $K'$ valleys, $s=\pm1/2$ to spin up and down states, and $g_v$ and $g_e=2$ are the valley and spin $g$-factors, respectively. The $\tau s$ product is the same for all the upper valence band LL states, e.g. $\tau s = 1/2$.  If the ratio between the Zeeman and cyclotron LL splitting of $(1+g_v)m^{*}/m_0$ is close to an integer, the QHSs revert to an even filling factor sequence at low $B$-fields, and a full lifting of the LL degeneracy at high $B$-fields \cite{Chu_PRB2014}. Theoretically, $g_{v}=2+\alpha$, where the first term stems from $d$-orbital magnetic moment, and the second is associated with the valley Berry phase \cite{DiXiao_PRL2007}. The measured $m^{*}$ combined with the theoretical expression for $\alpha=m_0/m^{*}$ \cite{DiXiao_PRL2007} yields a ratio of Zeeman to cyclotron energy of 2.3. Recent magneto-optical studies in monolayer WSe$_2$ \cite{srivastava2015valley,aivazian2015magnetic,Mitioglu_NL2015} confirm the $d$-orbital contribution to the valley $g$-factor, although the reported $g_v$ values differ, and individual LLs were not resolved.

Resolving individual layer densities in bilayer WSe$_2$ indicates that the layers are weakly coupled, which contrasts the more familiar case of Bernal stacked bilayer graphene, where the strong van der Waals interlayer coupling ($\sim$ 0.4 eV) significantly alters the energy momentum dispersion compared to monolayer graphene.  Calculated values of the interlayer coupling in bilayer WSe$_2$ range from 60 meV \cite{fang_textitab-initio_2015} to 67 meV \cite{gong_magnetoelectric_2013}. We can estimate an upper bound of the interlayer coupling for the WSe$_2$ samples examined here.  The minimum layer density difference of $3.5\times10^{12}$ cm$^{-2}$ determined from Fig.~\ref{fig3}(d) data, combined with a density of states $\frac{m^*}{\pi\hbar^2}=1.87\times10^{14}$ cm$^{-2}$eV$^{-1}$ yields a subband separation of 19 meV.

To summarize, we present a magnetotransport study of mono and bilayer WSe$_2$. The data reveals SdH oscillations and QHSs in both mono and bilayer samples that occur primarily at even filling factors.  The FT analysis evinces two subbands in bilayer WSe$_2$, located in the top and bottom layers, and negative compressibility of carriers in individual WSe$_2$ layers. We determine a hole effective mass of $0.45m_{0}$ in both mono and bilayer WSe$_2$.

% Acknowledgements
\begin{acknowledgments}
The authors acknowledge illuminating discussions with Xiao Li, and Qian Niu, and support from Intel Corp. and NRI SWAN. BF and HCPM contributed equally to this study. A portion of this work was performed at the National High Magnetic Field Laboratory, which is supported by National Science Foundation Cooperative Agreement No. DMR-1157490, and the State of Florida.
\end{acknowledgments}

%\bibliography{SdH_WSe2_Babak_v2}

%PARSED REFERENCE DATA

\end{document}